\documentclass{aa}
\usepackage{apj2aa}
\usepackage{astron} 
\usepackage{float,epsfig,psfig}
\usepackage{pstricks}

\newcommand{\mincir}{\raise
  -2.truept\hbox{\rlap{\hbox{$\sim$}}\raise5.truept \hbox{$<$}\ }}
\newcommand{\magcir}{\raise
  -2.truept\hbox{\rlap{\hbox{$\sim$}}\raise5.truept \hbox{$>$}\ }}

\begin{document}

\title{XMM-Newton discovery of an X-ray filament in Coma.} 
\author{A. Finoguenov\inst{1}, U. G. Briel\inst{1}, J.P.Henry\inst{2}}

\offprints{A. Finoguenov, alexis@xray.mpe.mpg.de}

\institute{Max-Planck-Institut f\"ur extraterrestrische Physik,
             Giessenbachstra\ss e, 85748 Garching, Germany 
\and
 Institute for Astronomy, University of Hawaii, 2680 Woodlawn Drive,
  Honolulu, Hawaii 96822, USA}

\date{Received January 12 2003; accepted August 12 2003}
\authorrunning{Finoguenov et al.}

\abstract{XMM-Newton observations of the outskirts of the Coma cluster
of galaxies confirm the existence of a soft X-ray excess claimed
previously and show it comes from warm thermal emission. Our data
provide a robust estimate of its temperature ($\sim0.2$ keV) and
oxygen abundance ($\sim0.1$ solar). Using a combination of XMM-Newton
and ROSAT All-Sky Survey data, we rule out a Galactic origin of the
soft X-ray emission. Associating this emission with a 20 Mpc region
in front of Coma, seen in the skewness of its galaxy velocity
distribution, yields an estimate of the density of the warm gas of
$\sim 50 f_{\rm baryon} \rho_{\rm critical}$, where $f_{\rm baryon}$ is the baryon
fraction of the gas and $\rho_{\rm critical}$ is the critical density
needed to halt the expansion of the universe. Our measurement of the
gas mass associated with the warm emission strongly support its
nonvirialized nature, suggesting that we are observing the warm-hot
intergalactic medium (WHIM). Our measurements provide a direct
estimate of the O, Ne and Fe abundance of the WHIM. Differences with
the reported Ne/O ratio for some OVI absorbers hints at a different
origin of the OVI absorbers and the Coma filament. We argue that the
Coma filament has likely been preheated, but at a substantially lower
level compared to what is seen in the outskirts of groups.  The
thermodynamic state of the gas in the Coma filament reduces the
star-formation rate in the embedded spiral galaxies, providing an
explanation for the presence of passive spirals observed in this and
other clusters.
 }

\maketitle
\section{Introduction}

The total amount of baryons directly observed in the local Universe is only
$\sim35$\% of the value implied from primordial nucleosynthesis, with that
observed at high-redshift (Fukugita et al. 1998, Table 3 excluding
line $7'$) and that observed within local closed box systems, such as
clusters of galaxies (Briel et al. 1992). Numerical
hydrodynamic simulations (e.g. Cen \& Ostriker, 1999; Dav\'e et al., 2001)
indicate that about half of these missing baryons reside in a warm-hot
intergalactic medium (WHIM). The temperature of the WHIM is expected to be
between $10^5$ and $10^7$ K with an overdensity with respect to the mean of
$\delta \sim 20$.  Most of the WHIM baryons are expected to reside outside
virialized regions in diffuse large-scale filaments.

\begin{table*}[ht]
{
\begin{center}
\footnotesize
{\renewcommand{\arraystretch}{0.9}\renewcommand{\tabcolsep}{0.12cm}
\caption{\footnotesize
Characteristics of the warm emission around Coma
\label{t:warm}}

\begin{tabular}{ccrrccccc}
 \hline
Coma & Exposure & $kT_{\rm Coma}$ & $kT_{\rm warm}$ & $Z/Z_{\odot}^{\sharp}$ &
$\delta_{\rm crit}^\flat$ & 0.5--2$^\natural$ &  \multicolumn{2}{c}{O$^\natural$}\\
field & ksec & keV & keV &&& keV  & VII &  VIII \\
\hline
0 &14.7&$15\pm6$  &$0.19\pm$0.01&$0.06\pm$0.02&$67\pm$7  &$10.3\pm$ 3.4 & $3.9\pm$1.4 & $1.6\pm$0.6\\ 
3 &11.9&$5.8\pm0.9$&$0.19\pm$0.02&$0.04\pm$0.02&$76\pm$12 &$ 8.6\pm$ 4.3 & $3.0\pm$1.6 & $1.2\pm$0.6\\
7 &21.2&$10\pm3$  &$0.22\pm$0.02&$0.07\pm$0.03&$45\pm$18 &$ 4.7\pm$ 2.0 & $1.2\pm$0.7 & $0.8\pm$0.5\\
11&13.3&$16\pm2$  &$0.24\pm$0.01&$0.09\pm$0.01&$71\pm$5  &$27\pm$ 3.0 & $5.4\pm$0.7 & $5.2\pm$0.7\\
13&22.4&$3.1\pm0.5$&$0.17\pm$0.01&$0.12\pm$0.04&$65\pm$7&$ 8.7\pm$ 2.9 & $4.7\pm$1.7 & $1.2\pm$0.4\\ 
\hline 
\end{tabular}
\begin{enumerate}
\item[$^\flat$]{\footnotesize ~ Filament overdensity relative to the critical,
    $(\rho/\rho_{\rm crit}-1)$, estimated by assuming a 90 Mpc distance, projected
    length of 20 Mpc, area of the extraction regions in Fig.~\ref{f:imh} and
    $n_e=8\times 10^{-7}$ cm$^{-3}$ in correspondence to $\rho_{\rm crit}$ for
    h=0.7} 

\item[$^{\sharp}$]{\footnotesize ~ Assuming a solar abundance ratio.}

\item[$^{\natural}$]{\footnotesize ~ flux, $10^{-7}$ photons cm$^{-2}$ s$^{-1}$ arcmin$^{-2}$}

\end{enumerate}
}
\end{center}
}
\vspace*{-1.cm}
\end{table*}

Several surveys have been undertaken to find the WHIM in absorption.
Tripp et al. (2000) have claimed a substantial fraction of the missing
baryons in the form of OVI absorbers. However, their observations lack
information on the ionization equilibrium as well as the O abundance
required to link the abundance of OVI ions to underlying O mass and then to
the amount of baryons. Chandra and XMM grating results on detection of OVII
and OVIII absorption lines (Nicastro et al. 2002; Fang et al. 2002;
Mathur et al. 2003; Fang et al. 2003; Rasmussen et al. 2003) allowed
for the first time an estimate of the ionization 
state of the gas, suggesting that the OVI absorbers account for over 80\% of
the local baryons. Still, such claims rely on the assumed O abundance.

Searches for the WHIM in emission with ROSAT have found some association of
soft X-rays with galactic filaments (Wang et al. 1997; Kull \& B\"ohringer
1999; Scharf et al. 2000; Zappacosta et al. 2002).  Upper limits have been
placed on emission by material connecting rich clusters of galaxies (Briel
\& Henry 1995) using stacked analysis of ROSAT All-Sky Survey data.  The
corresponding limit on the emitting gas density is $\delta < 375$, assuming
a temperature of 1 keV and an iron abundance of 0.3 solar. These
observations may also be used to constrain the WHIM because most of it is
expected to be in filaments connecting clusters.

The low filament densities implied by the weak absorption and soft X-ray
emission, and expected from typical overdensities in simulations,
complicates WHIM detection and study, unless the filament containing it is
viewed along the line of sight. A combination of both the requirement that a
filament be so viewed and the expectation that clusters are at the
intersection of filaments, indicates that searches for WHIM near clusters of
galaxies may prove fruitful.  The soft X-ray excess above the harder
intracluster medium (ICM) emission reported for some clusters (Lieu et
al. 1996; Bonamente et al. 2002) may be signaling a fortunate orientation of
a filament containing WHIM at those clusters.

%\begin{figure*}
\includegraphics[width=8.5cm]{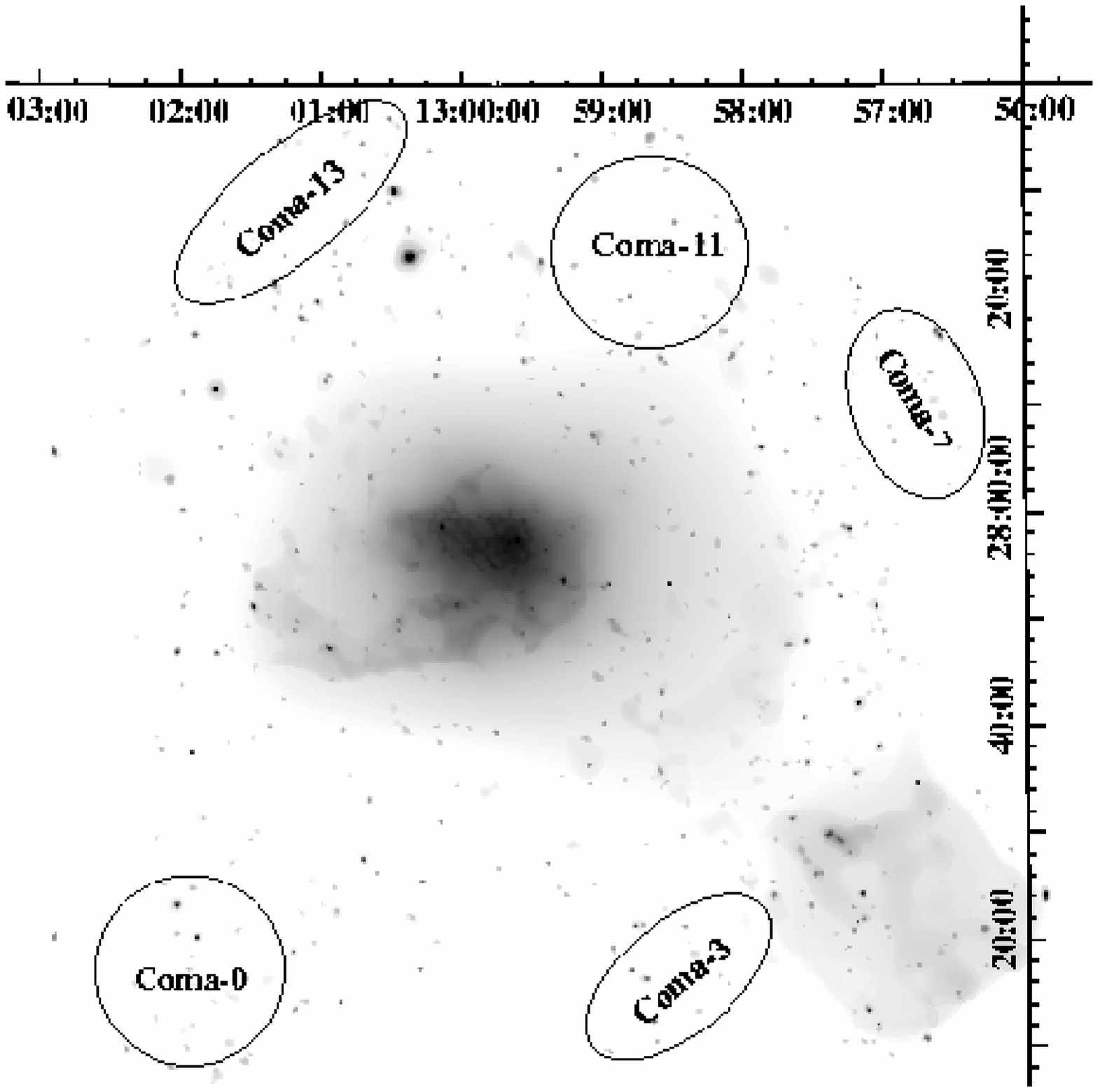}
%plots/coma_regs.ps}

\figcaption{An image of the Coma cluster in the 0.5--2 keV band, showing the
positions of the spectral extraction regions (solid ellipses) with names
according to the XMM-Newton survey notation. The coordinate grid marks R.A.,
Dec. (J2000.0).
\label{f:imh}
}
%\end{figure*}

In this Paper we carry out an X-ray spectroscopic study of the
outskirts of the Coma cluster, which has a soft excess, in order to search
for and characterize any WHIM at this location. Our observations are from
the mosaic made by XMM-Newton. Kaastra et al. (2003) report a similar study
of the center of the Coma cluster also using data from the XMM mosaic. Not
knowing the relative spatial distributions of the WHIM and ICM, it is
difficult to predict whether observing the center or outskirts of a cluster
will maximize the contrast of any warm component relative to the dominant
hot component. However both studies do detect a warm component. We adopt a
Coma cluster redshift of 0.023, $H_{\rm o}=70$~km~s$^{-1}$~Mpc$^{-1}$, and
quote errorbars at the 68\% confidence level. One degree corresponds to 1.67
Mpc.

\section{Observations}\label{s:res}

The initial results of the XMM-Newton (Jansen et al. 2001) performance
verification observations of the Coma cluster are reported in Briel et al.
(2001), Arnaud et al. (2001) and Neumann et al. (2001). In addition to the
observations reported there, three additional observations have been carried
out, completing the planned survey of the Coma cluster. We have renamed the
Coma-bkgd field from Briel et al. (2001) Coma-0 in order to avoid confusion
with the additional background pointings required to analyze these data.

Since the Coma cluster is a large diffuse source, estimating the
background is non-trivial. In general we follow the method of Lumb et
al. (2002). The observations reported here are performed with the EPIC pn
detector (Str\"uder et al. 2001) using its medium filter. This configuration
leads to a different detector count rate from the cosmic (CXB) and Milky Way
X-ray backgrounds in the soft band, compared to the usual thin filter. The
Milky Way background is further composed of the Local Bubble and halo
components. We used a 100 ksec observation of the point source APM08279+5255
(PI G. Hasinger), performed with the same (medium) filter, to measure these
background components and a filter wheel closed observation to measure
the internal or particle background. We used the APEC plasma code
(Smith et al. 2001) to fit the Milky Way background components and a
power-law ($\Gamma=1.45$) to fit the CXB. The element abundance in the APEC
code was fixed to solar and temperature was set to 0.07 keV for the Local
Bubble and left as a free parameter for the halo, yielding a 0.22 keV
best-fit value. To ensure the compatibility of the CXB, we excised
from the spectral extraction region the APM quasar and the X-ray sources
identified with Coma galaxies (which otherwise increase the apparent CXB
flux by 10\%). Since there is a mismatch in the $N_H$ toward the Coma and
the APM fields, we use the same spectral shape and normalization of the halo
and CXB components, but changed their $N_H$ from $3.9\pm0.3\times10^{20}$
cm$^{-2}$ (APM) to $9\times10^{19}$ cm$^{-2}$ (Coma). The Local Bubble is
not subject to this absorption.  We also checked that the derived results on
the soft X-ray emission from Coma do not depend on the possible large-scale
variation in the intensity of the Local Bubble component.

For the spectral analysis we use the 0.4--6 keV range and for the
estimate of the detector background the 9.5--15 keV range. We do not
use the energies below 0.4 keV, to avoid a complicated treatment of
detector background associated with the electronic noise (Lumb et
al. 2002). This precludes us from studying the very soft
emission. Although, we could potentially use the energies up to 10 keV
in the analysis, we are able to sufficiently constrain the hot
component of the Coma cluster already without using these energy
bands, where removal of the detector background becomes critical. 

The vignetting correction is performed taking the source extent and a
recent pn telescope (XRT3) vignetting calibration (Lumb et al. 2003)
into account, which is mostly important for the absolute flux
determination, given our choice of the energy range. Systematic
uncertainty of the flux determination is 4\% for both pn and MOS. 

In addition to a thorough evaluation of the background, we
further reduce the systematics of our study by using the results of
recent calibration efforts of the pn team of MPE on soft energy
response of the pn detector (Haberl et al. 2002, XMMSAS-5.4
release). Furthermore, to improve on the detection statistics of the
warm emission and to reduce the systematic effects of subtraction of
the cluster emission, we concentrate on the {\it outskirts} of the
Coma cluster.

We investigated some systematic errors in our analysis by determining the
effects of using experimental software or non-standard values of some
parameters.  We do not use these modifications in our analysis, however. The
effect of pn energy scale calibration was evaluated by comparing the results
of the standard analysis with experimental software that corrects the zero
point of the energy scale for each pixel of the pn. We found some subtle
differences, e.g. a higher best-fit temperature by 0.03 keV, which is larger
than the statistical error quoted in Table \ref{t:warm}. However, the ratio of
O to Ne was still found to be solar or larger (see Sect. 5). Inclusion of
ROSAT All-Sky Survey (RASS) data into the analysis of the background allowed
us to constrain better the Local Bubble component, e.g. a temperature of
$0.10\pm0.01$ keV was found but with a somewhat lower luminosity. Using
these values instead of the standard ones quoted above had no effect on our
results.

In Fig.~\ref{f:imh} we show the location of the spectral extracting
areas.  We select regions $\sim40'$ from the Coma center to the
North-West, North, North-East, South and South-East. The South-West
direction is complicated by the presence of the infalling subcluster
(NGC4839), while some other pointings were affected by background
flares, which after screening lead to insufficient exposures.

\section{Results}

Our major results are shown in Fig.~\ref{f:spepn} and listed in Table
\ref{t:warm}. We find that the soft excess in Coma on spatial scales of
$30'-50'$ is characterized by thermal plasma emission, because of the clear
detection of O lines, with a characteristic temperature of $\sim0.2$
keV. Bonamente et al. (2003) found the same temperature
characterizing the ROSAT PSPC observation of the Coma soft excess. The
highest temperature of this component corresponds to the hot spot in Coma
found in ASCA observations (Donnelly et al. 1999), and identified as an
accreting zone of the Coma cluster.

We have performed several internal checks of our results.  First,
Fig.~\ref{f:spem1} shows the confirmation of our findings for the Coma-11
field by analysis of the XMM-Newton MOS1 data. The medium filter was also
used for the MOS during these observations. For this analysis we have done a
simple background analysis that uses the flux in the MOS1 field for the same
instrument settings and background conditions but a different $N_H$. Since
the MOS temperature of the hot component is consistent with that determined
by the pn, we fix the temperature of that component to its pn value in the
MOS spectral analysis.  The MOS fit then agrees with the pn results for the
intensities of the hot and warm components and for the temperature and
element abundance for the warm component.

Second, to entirely exclude the effect of the instrumental or particle
background, we have attempted an in-field background subtraction for the
pn. Using the difference in the spatial distribution of the Coma hot and
detector background components within the Coma-11 field, we were able to
remove both of them {\it using data only from the Coma-11 field}. In
particular, the region south of the pointing center contains more of the
cluster emission relative to the detector background and vice-versa for a
region north of the pointing center. Since the spectrum of the CXB is
similar to that of Coma hot emission (see Fig.~\ref{f:spepn}), it could be
removed by over-subtracting the Coma hot flux. We exclude energies
corresponding to the detector emission lines from the comparison, as their
distribution is different from that of the continuum (e.g. Lumb et
al. 2002). The resulting spectrum is displayed in Fig.~\ref{f:spepn} and
could be fitted with a single MEKAL model with $0.22\pm0.01$ keV temperature
and a factor of 3 higher intensity than the Milky Way halo plus
Local Bubble emission. Third, we have tried several compilations of the XMM
background (Lumb et al. 2002; Read 2003) and found that the soft component
in the Coma emission is clearly seen using any of them.

\includegraphics[width=8.cm]{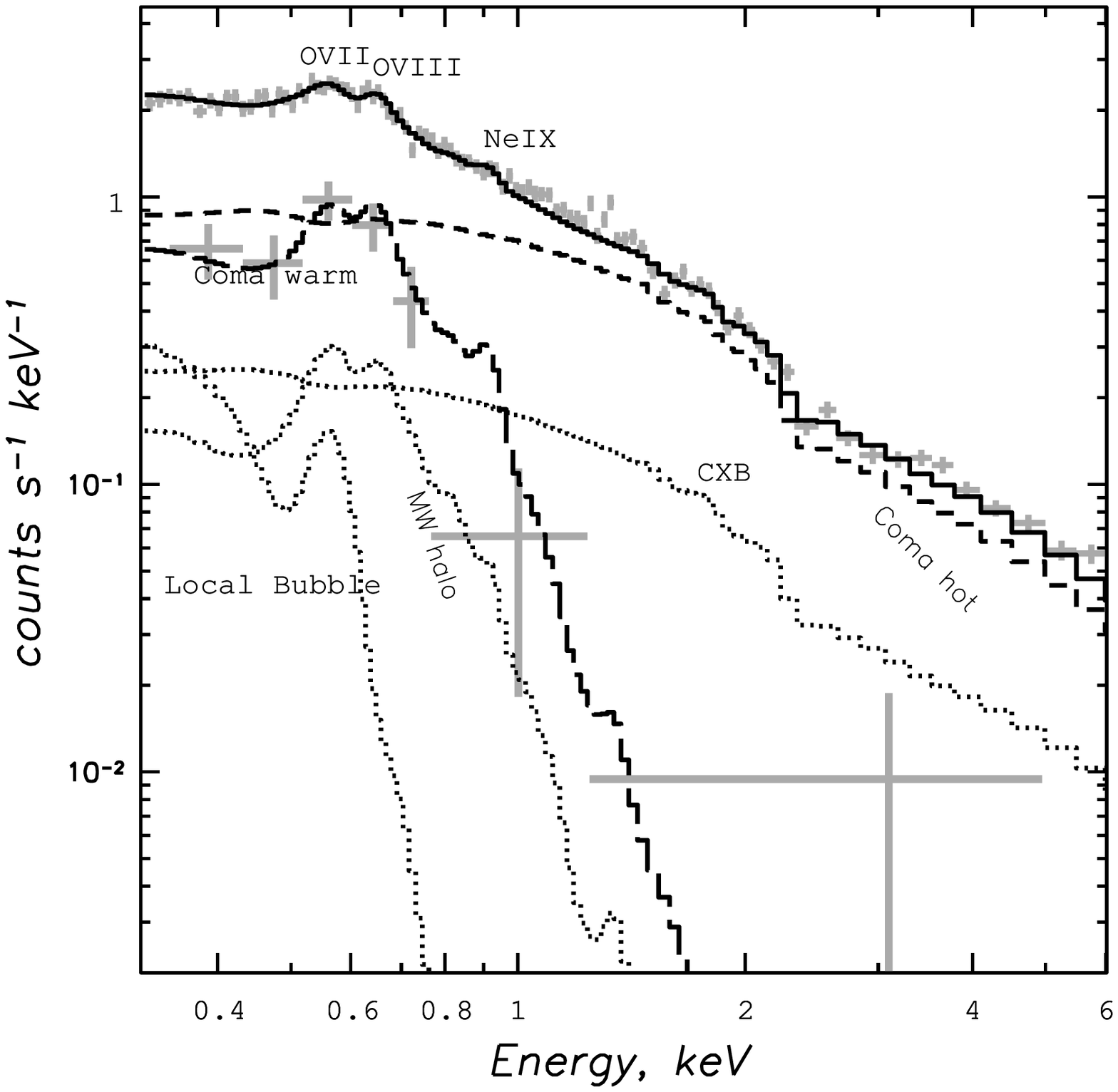}
%{plots/spe_c11_apm2.ps}

\figcaption{Detailed decomposition of the pn spectrum (small grey crosses)
in the Coma-11 field into foreground and background components, obtained
from the analysis of the APM field, plus hot and warm emission from the Coma
cluster. Large crosses are the result of the {\it in-field} subtraction of
Coma emission and detector background, possible due to differences in the
spatial distributions between the Coma warm, Coma hot and detector
background components. MW halo means Milky Way halo and CXB means cosmic
X-ray background.
\label{f:spepn}
}

\includegraphics[width=8.cm]{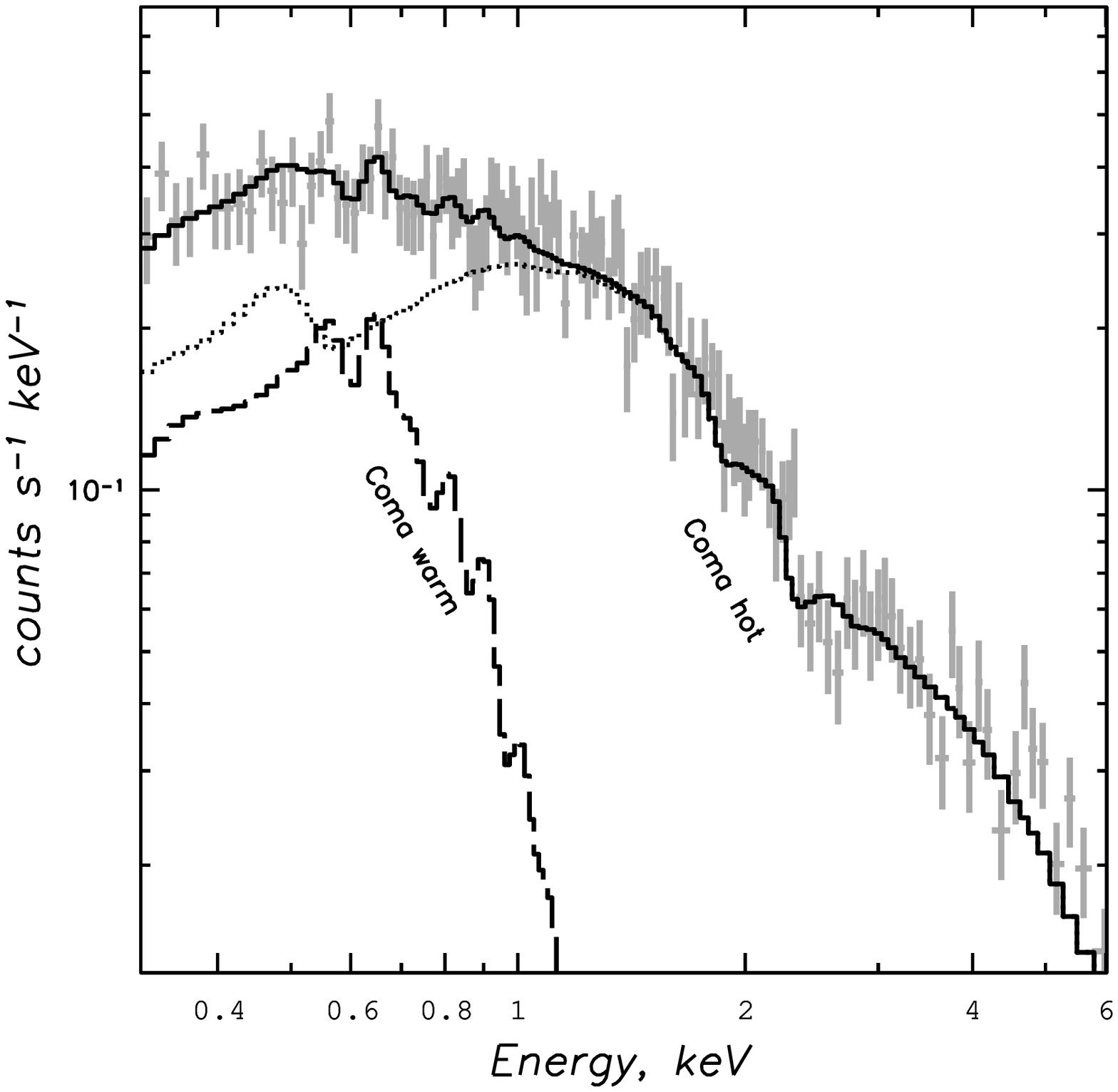}
%{plots/spe_c11m1_apm.ps}

\figcaption{MOS 1 APM field background subtracted spectrum of the Coma-11
field. Detailed modeling, accounting for the differences in $N_H$ was not
performed, but the difference is small, given the lower sensitivity of MOS
CCDs at low energies compared to the pn.
\label{f:spem1}
}

At the moment, it is difficult from the O line redshift alone to
decide whether this emission is Galactic or extragalactic. However,
some apparent differences with the Galactic emission are already
noticeable. The soft component is centered on Coma and has an
amplitude exceeding the variation of the underlying Galactic emission
(Bonamente et al. 2003). An O abundance of 0.1 times solar is much
lower than the Galactic value of one solar (Markevitch et al. 2002;
Freyberg \& Breitschwerdt 2002). The temperature variations of the
Coma warm component are not seen in the quiet zones of Galactic
emission at high latitudes.  The observed warm component exceeds the
level of the Milky Way halo plus Local Bubble continuum (at 0.57 keV)
emission by a factor of 3. Also the strength of the observed O lines
are a factor of three larger than the corresponding values for the
Galactic emission, and to significantly change the derived O
abundance of the Coma warm emission a possible variation in the metallicity
for the Galactic emission should be by a factor of $\sim 3$, which is not
observed (Freyberg \& Breitschwerdt 2002). As we will demonstrate
below, ROSAT PSPC data strongly constrain the contribution of the
low-temperature Galactic components to the observed O line flux.

\includegraphics[width=8.cm]{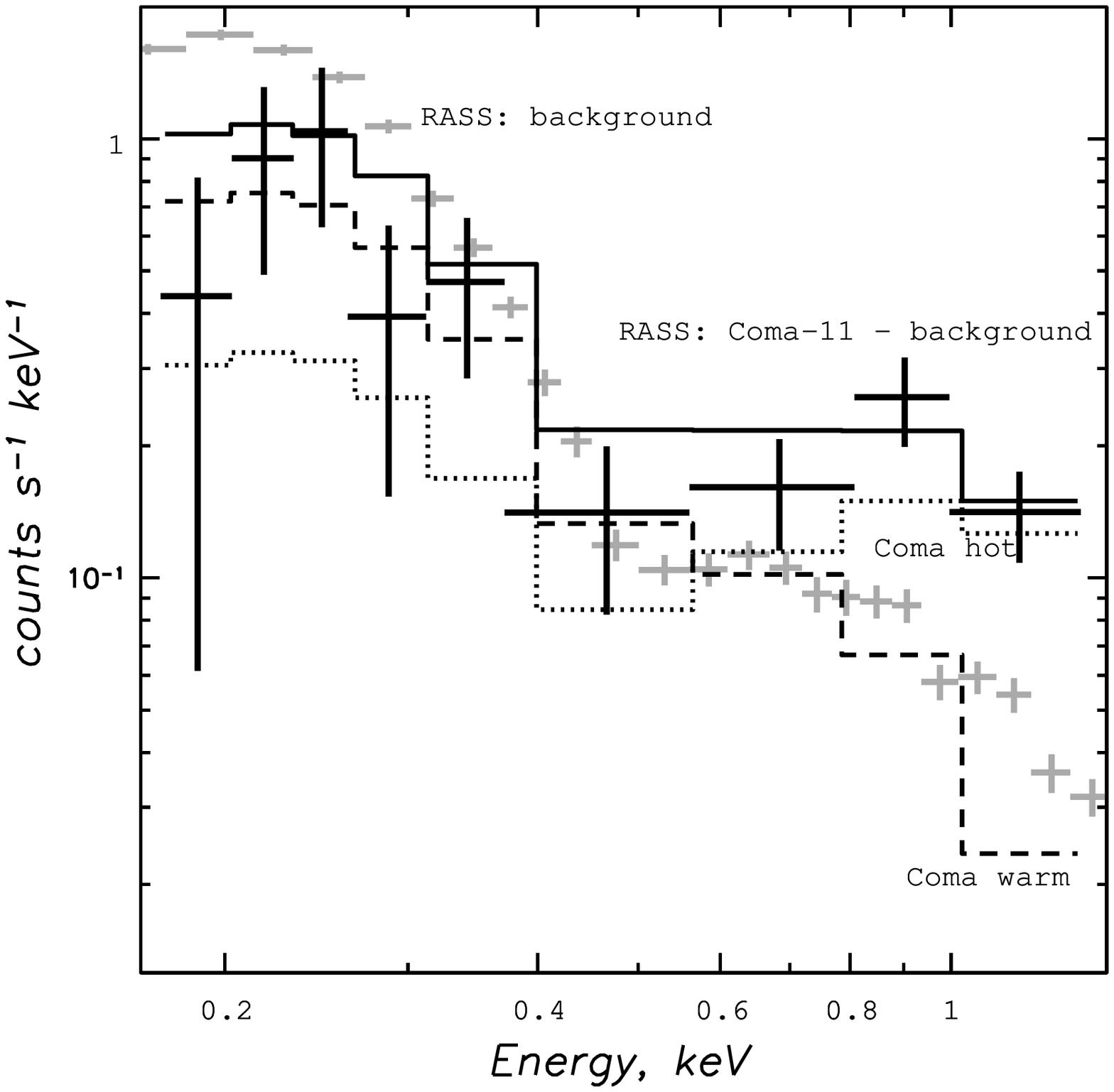}
%{plots/spe_c11_rass.ps}

\figcaption{ROSAT All-Sky Survey spectrum of Coma-11. Grey crosses show the
background spectrum $2^\circ$ north of the Coma center.  Black crosses show
the background subtracted spectrum extracted at the position of the Coma-11
field. Lines show the best fit spectrum from the analysis of pn data folded
through the ROSAT PSPC with no adjustment to the pn best fit. Dashed line is
the warm component, dotted line is the hot component and the solid line is
the sum. The best fit model to the pn data provides a statistically
acceptable fit to the RASS data ($\chi^2=9.8$ per 9 d.o.f.).
\label{f:rass}
}

\section{Comparison with the ROSAT All-Sky Survey}

We use the RASS to provide an external check of our results. We note,
however, that the spatial and spectral resolution and energy band of
the ROSAT detectors are significantly different from those on XMM so
care needs to be taken when performing this check.

We first compare the fluxes implied by our analysis to those in the
RASS data as presented in Snowden et al. (1997). We choose the
0.5--0.7 keV band (ROSAT R4 band), corresponding to position of the O
lines and the 1.3--2.0 keV band (R7), where the relative contribution
of the hot component is the largest. We make the comparison for two
fields: Coma-11 and a field $4^\circ$ North of Coma, which we call Sky
Background. Several effects should be taken into account to provide a
proper comparison between the XMM-Newton and ROSAT data. First, the
X-ray mirrors of XMM-Newton have a 7\% efficiency to stray-light (Lumb
et al. 2002) with uncertainty in the energy dependence of 2\%. This
results in higher fluxes seen by the XMM detectors near bright objects
and may lead to object-specific flux variations. We use the RASS data
to estimate the effect of stray-light pollution in the background and
Coma fields by assuming uniform azimuthal and radial dependence of the
stray-light coming from XMM off-axis angles of
$0.4^\circ-1.4^\circ$. This component should be subtracted from the
XMM fluxes in the comparison. Second, the RASS maps of the diffuse
X-ray background have point sources removed, which reduces the
intensity of the background. We use the limits of Snowden et
al. (1997) and the $log N -log S$ of Hasinger et al. (2001) to
estimate this removed flux. This component should be added to the RASS
flux in the comparison. Third, RASS data on the sky background on the
scales of the XMM field of view have rather large statistical
uncertainties and some averaging should be carried out in order to
provide a meaningful comparison. Fourth, the R4 images exhibit
variations larger than the statistical errors, so we quote the errors
corresponding to the systematic variations. Fifth, in the analysis of
XMM spectra a separate power-law component was introduced to remove
the increased detector background due to soft protons (De Luca \&
Molendi 2002). Studies of this component by several groups (Katayama
et al. 2002; Snowden S., Molendi S. 2003, private communications) show
that its intensity could be determined by the spectrum at high
energies (10--15 keV) where the mirrors have no collecting power for
X-rays, while the instrumental background originates from energy
losses of particles on the detector. From the comparison of the XMM
predicted and the RASS observed intensities of the Sky Background
presented in Table \ref{t:bkg} it is clear that our background model is
applicable for the Coma cluster. Our results for Coma-11 are also in
agreement with the RASS flux, although the later has a rather large
uncertainty.

The pixel size in the Snowden maps is rather large ($12^\prime$), so
we have performed a second analysis by extracting a spectrum from the
raw RASS data at the exact location of the Coma--11 field, a circle of
$8^\prime$ radius centered on $\alpha=194.66521^\circ$,
$\delta=28.406924^\circ$. We have extracted a background spectrum
using a $0.6^\circ \times 2^\circ$ region (long axis perpendicular to
radius vector from the cluster center) $2^\circ$ North of the
cluster. We display the results in Fig.~\ref{f:rass}. We find that our
pn model provides an acceptable fit to the RASS data with no
adjustable parameters ($\chi^2=9.8$ per 9 degrees of freedom). 

Fig.~\ref{f:rass} also shows that a simple comparison of the countrates
between the two instruments is not straightforward, particularly at energies
below 0.5 keV. The ROSAT background is very high in this energy range
compared to the pn. This is caused by two differences between the
instruments. At an energy of 0.25 keV the effective area of the ROSAT PSPC
is about a factor of two higher than the XMM pn ($\sim175$ cm$^{-2}$
vs. $\sim80$ cm$^{-2}$) while the energy resolution of the PSPC is about a
factor of two worse than the pn ($\sim1$ vs. $\sim2$) (Snowden et al. 1997
for the PSPC, XMM Users Handbook for the pn). At 0.25 keV all components are
dominated by the Local Bubble. The poor energy resolution of the ROSAT PSPC
redistributes the larger number of counts from this component to higher
energies. On the other hand, since the XMM-Newton Coma observations were
made with the medium filter, most of the counts at lower energies are
actually redistributed from higher energies. This is why our results are not
so sensitive to the model of the Local Bubble. The combination of these two
effects produces substantial differences in the ratio of the Coma counts to
the background counts below 0.5 keV in the two instruments and a proper
comparison requires detailed spectral modeling as Fig.~\ref{f:rass} shows.

The comparison of Fig.~\ref{f:spepn} and Fig.~\ref{f:rass} reinforces our
previous analysis using the Snowden maps, but with a finer spatial and
spectral resolution. For example, the intensity of the total background
continuum at 0.57 keV in both figures is about equal to the Coma warm
continuum. However, the unlimited field of view of the RASS allows a direct
measurement of the background in Fig.~\ref{f:rass} compared to a model of the
background in Fig.~\ref{f:spepn}. Thus we conclude that an incorrect
background model producing the observed Coma Warm component is ruled out.

Our analysis of the ROSAT data can also be used to set limits on the neutral
gas fraction in a filament. From Table 1, the HII column is about $10^{21}$
cm$^{-2}$. From Fig.~\ref{f:rass} the limit on the column of HI above the
Galactic value is $<10^{20}$ cm$^{-2}$. Hence, the HI fraction is less than
10\%. Marginal indication of such absorption is present in Fig.~\ref{f:rass},
as the data points are slightly below the model prediction.

\begin{table}[!t]
{
%\begin{center}
\footnotesize
{\renewcommand{\arraystretch}{0.9}\renewcommand{\tabcolsep}{0.09cm}
\caption{\footnotesize
Comparison between XMM-Newton Predictions of the ROSAT Surface
Brightness from Observations (Coma-11) or Background Model (Sky Back)
and Observed ROSAT All-Sky Survey Data.
\label{t:bkg}}

\begin{tabular}{ccccc}
\hline
Field & \multicolumn{2}{c}{XMM-Newton pn}  & \multicolumn{2}{c}{RASS} \\
name  & predicted & stray-light & data & removed flux \\
     &  \multicolumn{4}{c}{$10^{-4}$ \ PSPC-C counts \ s$^{-1}$ \ arcmin$^{-2}$} \\
\hline
\multicolumn{4}{c}{0.5-0.7 keV}\\
Coma-11 & $2.7\pm0.1$   & 0.10 & $2.0\pm0.5$ & 0.2 \\  
Sky Back & $0.66\pm0.08$ & 0.05 & $0.5\pm0.1$ & 0.2 \\
\multicolumn{4}{c}{1.3-2.0 keV}\\
Coma-11 & $3.2\pm0.1$   & 0.14 & $3.2\pm0.7$ & 0.2 \\  
Sky Back & $0.56\pm0.04$ & 0.05 & $0.4\pm0.02$ & 0.2 \\
\hline 
\end{tabular}
}}
%\vspace*{-1.cm}
\end{table}

\section{Interpretation}

In the rest of this Paper we present an interpretation assuming an
extragalactic origin of the warm component.  With the CCD-type spectrometry
we can place limits on the possible redshifts of the warm component, which
for the Coma-11 field is determined as $0.007\pm0.004\pm0.015$ (best-fit,
statistical and systematic error). Any extragalactic interpretation should
therefore concentrate on the large-scale structure in front of the Coma
cluster since the allowed redshift is 0.000 - 0.026 compared to the Coma
value of 0.023.  We have carried out an analysis of the CfA2 galaxy catalog
(Huchra et al. 1995) towards the Coma cluster, excising the
South-West quadrant, where strong influence of the infalling subcluster
NGC4839 on the velocity distribution has been suggested (Colles \& Dunn
1996). Fig.~\ref{f:ghst} illustrates our result, indicating a significant
galaxy concentration in front of the Coma cluster, with velocities lying in
the $4500-6000$ km s$^{-1}$ range. The sky density of the infall zone is
shown in the insert of Fig.~\ref{f:ghst}. It was constructed from the number
of galaxies in the $4500-6000$ km s$^{-1}$ velocity range minus the number
of galaxies in the $8500-10000$ km s$^{-1}$ bin, corresponding to a similar
difference in the velocity dispersion from the cluster mean. By excising the
galaxies at velocities lower than 6000 km s$^{-1}$, we recover a projected
velocity dispersion of $\sim750$ km s$^{-1}$. Using the $M-\sigma$ relation
of Finoguenov et al. (2001), we find that such velocity
dispersion is more in accordance with mass of the Coma cluster. The excess
of galaxies exhibits a flat behavior within 0.8 degrees (1.3 Mpc), followed
by a decline by a factor of 5 within 1.7 degrees (2.8 Mpc) and a subsequent
drop by two orders of magnitude within 10 degrees (16.7 Mpc) The soft
emission from the Coma is detected to a 2.6 Mpc distance from the center
(Bonamente et al. 2003) in a remarkable correspondence to the galaxy
filament.

In the absence of the finger--of--god effect (decoupling of galaxies
from the Hubble flow), validated below by our conclusion on the
filamentary origin of this galaxy concentration, the infall zone is
characterized by a 20 Mpc projected length. This length assumption
affects the density estimates, while the O abundance is only based on
the assumption of collisional equilibrium. We have verified the later
assumption by studying the ionization curves for OVII and OVIII
presented in Mathur et al. (2003). We have concluded that the
assumption of pure collisional ionization is valid for our data, since
$n_e>10^{-5}$ cm$^{-3}$ and $T>2\times10^{6}$ K (0.17 keV). An
advantage of our measurement is that we also determine the temperature
by the continuum. Lower temperatures, which at densities near
$10^{-5}$ cm$^{-3}$ result in similar line ratios for OVII and OVIII,
fail to produce the observed (H-e and He-e) bremsstrahlung flux at
0.7--1 keV.

An important question we want to answer is whether the detected emission
originates from a group or from a filament. We cannot decide from the
electron density of the structure ($\sim50 (\mu_e / \mu_p) f_{\rm baryon}
\rho_{\rm crit}$), as it suits both (for the low metallicity of
filament we take 
$\mu_e=1.1$, $\mu_p=1.2$ and $f_{\rm baryon}=0.16$ to correspond to the first
WMAP results in Spergel et al. 2003). However, the implied mass of the
structure is comparable to the mass of the Coma cluster
(see also Bonamente et al. 2003). On the other hand the temperature of the
emission is $\sim0.2$ keV, almost two orders of magnitude lower than that of
the Coma cluster. It takes a few hundred groups with virial temperature of
0.2 keV to make up a mass of the structure, which leads to an overlapping
virial radii if we are to fit them into the given volume of the
structure. Thus we conclude that the observed structure is a filament.

\includegraphics[width=8.cm]{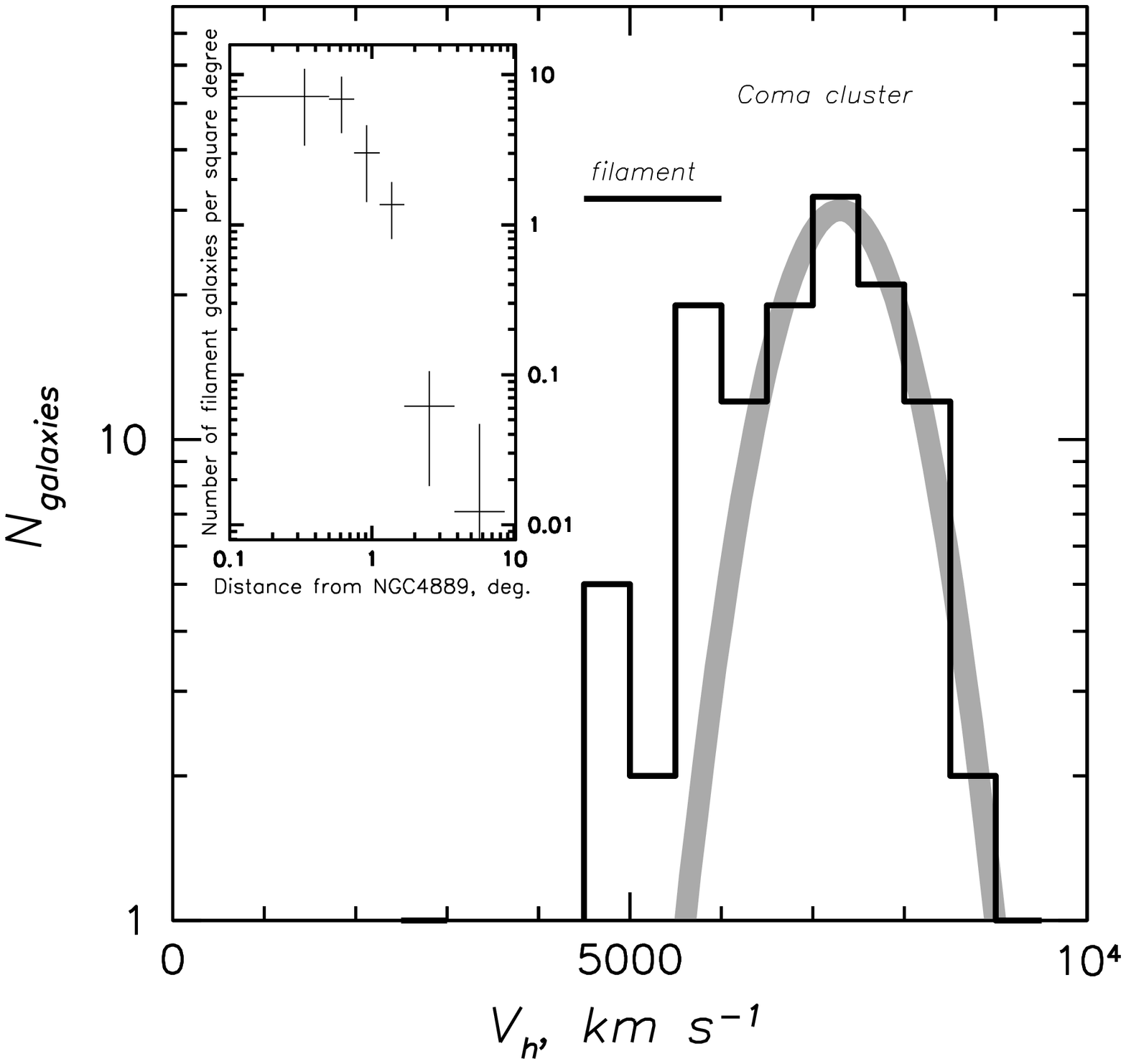}
%{plots/glx_hist4.ps}

\figcaption{Galaxy distribution in the direction to the Coma cluster from
CfA2 survey. Central 2 degrees in radius field centered on NGC4889 are shown,
excluding the 60 degrees cone in the South-West direction from the center,
to avoid the effect of the NGC4839 subcluster. An excess of galaxies over
the Gaussian approximation of the velocity dispersion of the Coma cluster is
seen in the $4500-6000$ km s$^{-1}$ velocity range. The insert shows the
spatial distribution of the excess galaxies in the $4500-6000$ km s$^{-1}$
velocity bin over the galaxies in the $8500-10000$ km s$^{-1}$ velocity bin,
this time out to 10 degrees.
\label{f:ghst}
}
\smallskip

For the Coma-11 field, where the statistics are the highest, we investigated
the effect of relaxing the assumption of solar abundance ratios. We note
that an assumption for C abundance is important for overall fitting and a
solar C value would affect our results. When left free, however, the C
abundance tends to go to zero. Also, there is no systematic dependence of
our results on the assumed C abundance as long as the C/O ratio is solar or
less, which is the correct assumption from the point of view of chemical
enrichment schemes and observations of metal-poor stars in our
Galaxy. Significant abundance measurements are: O/O$_{\odot}=0.14\pm0.02$,
Ne/Ne$_{\odot}=0.14\pm0.06$, Fe/Fe$_{\odot}=0.04^{+0.03}_{-0.01}$ (assuming
Fe$_{\odot}/H=4.26\times10^{-5}$ by number), indicating that Fe is
underabundant by a factor of three in respect to the solar Fe/O ratio,
implying a dominant contribution of SN II to Fe enrichment.  Element
abundances for the hot emission obtained from our XMM data on the center of
Coma are Mg/Mg$_{\odot}=0.36\pm0.12$, Si/Si$_{\odot}=0.45\pm0.08$,
S/S$_{\odot}=0.01\pm0.15$, Fe/Fe$_{\odot}=0.20\pm0.03$, also suggesting
prevalence of SN II in Fe enrichment, although Fe abundance is a factor of 5
higher compared to the filament. The filament Ne/O ratio reveals a subtle
difference with the OVI absorbers. Nicastro et al. (2002) reports Ne/O$=2$
times solar for their X-ray absorption observations of material associated
with the OVI absorbers. We observe a lower (solar) ratio at 95\%
confidence. Lower ratios seem to be a characteristic of enrichment at $z>2$
(Finoguenov et al. 2003b), while simulations suggest that the
WHIM originates at $z<1$ (Cen \& Ostriker 1999).

While we would rather have more observational evidence on the dispersion of
O to alpha element ratios at different sites, we believe that this is a
major source of systematics in interpreting the element abundance of X-ray
filaments as a universal value.
To illustrate the point, we calculate the density of baryons traced by
the OVI absorbers under two sets of assumptions.  Scaling the
original value of $\Omega_b(OVI) = 0.0043h_{70}^{-1}$ of Tripp et
al. (2000) for ionization equilibrium implied by measurements of
Mathur et al. (2003) and using the measurements of the O abundance
reported here yields:
\begin{equation}
\Omega_b^{OVI} = 0.020h_{70}^{-1} \left( {0.14 \over 10^{[O/H]}}\right) \left({\langle
[f(OVI)]^{-1}\rangle \over 32}\right)
\end{equation}
The formal errorbar is 0.006, mostly from the uncertainty in the
estimate for ionization. If, on the other hand, our measurements of Ne
abundance are used with the Ne/O ratio for OVI absorbers from Nicastro
et al. (2002) then,
\begin{equation}
\Omega_b^{OVI} = 0.040h_{70}^{-1} \left( {0.14 \over 10^{[Ne/H]}}
 { 10^{[Ne/O]}  \over 2}\right)
\left({\langle [f(OVI)]^{-1}\rangle \over 32}\right)
\end{equation}
Therefore the second set of assumptions must be invalid since the total
baryon density is $\Omega_b^{\rm total}=0.039$, leaving no room for other
major components of local baryons such as Ly$_\alpha$ absorbers
($0.012\pm0.002$) and stars and clusters of galaxies ($\sim0.006$; e.g.,
Finoguenov et al. 2003b and references therein). O depletion onto dust
grains in the OVI absorbers, as suggested by Nicastro et al. (2002) as an
explanation of the high Ne/O ratio, does not explain the unacceptably high
baryon abundance implied by Eq. (2), since the solar Ne/O ratio in our
observations may be simply explained by dust sputtering.

The observational determination of scaling relations between X-ray
properties, such as luminosity $L_X$, gas temperature $T$, and entropy
is crucial in establishing the physical properties of the ICM. The
slopes of the $L_X-T$ and mass--temperature relations (e.g.
Markevitch 1998; Finoguenov et al. 2001) and the gas
entropy level (e.g. Ponman et al. 1999; Finoguenov et
al. 2002) are all at variance with model predictions based on pure
gravitational heating (Kaiser 1986) and require the introduction of
extra physics to describe the thermodynamics of the ICM (e.g. Evrard
\& Henry 1991; Kaiser 1991).  An often discussed piece of extra
physics is preheating of the baryons before they accrete onto the
cluster. Most baryons that accrete onto clusters are thought to come
from filaments, so we now have an opportunity to compare the
thermodynamic properties of the filament gas with that of group and
cluster gas. The entropy of the X-ray emitting filament is 150 keV
cm$^2$, which could be reproduced by heating while falling onto a
filament (e.g. Cen \& Ostriker 1999). In fact, estimating the expected
Mach number of the accretion shock when the filament enters the Coma
cluster yields a value of 4, similar to predictions of simulations by
Miniati et al. (2000).  Regardless of the origin of the entropy of the
filamentary gas, its entropy is much smaller than the 400 keV cm$^2$
implied by ASCA observations of the outskirts of groups (Finoguenov et
al. 2002), ruling out a universal preheating value. We note that the
energy of the gas (${3\over2}kT=0.36\pm0.02$ keV/particle) is similar,
although higher than the SNe energy associated with enrichment of gas
to the observed O abundance ($0.22\pm0.03$ keV/particle), so heating
by galactic winds is not ruled out.

Although, the temperature of the filament is presently a factor of 40 lower
than the Coma virial temperature, as it falls into the cluster it will be
shock heated, and its final adiabat will be higher than in the self-similar
case (Dos Santos \& Dor\'e 2002; Ponman et al. 2003). Since
the mass of the filament is comparable to that of the Coma cluster, the
combined object will also deviate from cluster scaling
relations. Observations indicate that the gas-to-dark matter distribution
will not be affected (Sanderson et al. 2003), but the temperature is higher
for a given mass (Finoguenov et al. 2001). This process is probably not
universal, since some groups of galaxies have shallow gas profiles, thus
possibly pointing to adiabatic compression rather than shock heating of the
preheated gas, as proposed by Tozzi \& Norman (2001).

If the structures reported here are only associated with large
clusters of galaxies, their contribution to the baryon budget is
negligible (0.1-1\%). Of greater importance is to investigate the
accreting environment of the massive groups and poor clusters that
have a significant entry in the baryon budget (6\%). However, early
enrichment epoch, suggested by low Ne/O ratio, makes X-ray filaments a
substantial entry in the metal budget at high redshifts.

The thermodynamic state of the filamentary gas causes a global feedback
effect on the embedded galaxies by strangling the gas accretion
(Finoguenov et al. 2003a; Oh \& Benson 2003). The resulting star-formation
will proceed by consumption of the previously accumulated gas, in either
quiescent mode as in the disk, or merger-induced bursts leading to
formation of the spheroid (Somerville et al. 2001).
Galaxy mergers are frequent inside galaxy groups (Kodama et al. 2001), but
in a filament the infall of the gas will primarily be recorded in the
star-formation history of the disk (Kennicutt et al. 1994).  A
relevant observation could therefore shed light on the feedback epoch, which
is crucial in understanding the relation between the X-ray filaments and OVI
absorbers. The recent Sloan Digital Sky Survey (SDSS) discovery of passive
spirals, in the same filament in front of Coma (Goto et al. 2003), is
exactly what is expected from this strangulation process. The existence of
these passive spirals lends further support to the association of the soft
X-ray excess with the Coma filament. As passive spirals are starting to be
found in the outskirts of many clusters, the presence of the soft X-ray
excess might be a wide-spread phenomenon among the massive clusters, as
indicated by first results on XMM-Newton REFLEX-DXL survey (Zhang et
al. 2003).

\section{Conclusions}

We have discovered oxygen line emission associated with the outskirts of the
Coma cluster. These data show that the previously observed soft X-ray excess
in this cluster is due to thermal emission from a warm gas, possibly the
long-sought WHIM. Typical parameters characterizing the emitting material
are a temperature $\sim0.2$ keV, abundance $\sim0.1$ solar, and density
$\sim 50 f_{\rm baryon} \rho_{\rm critical}$.  The thermodynamical
state of the gas 
in the Coma filament reduces the star-formation rate of the embedded spiral
galaxies, providing an explanation to the presence of the passive spirals in
the SDSS survey in this region (Goto et al. 2003).

\begin{acknowledgements}

The paper is based on observations obtained with XMM-Newton, an ESA science
mission with instruments and contributions directly funded by ESA Member
States and the USA (NASA). The XMM-Newton project is supported by the
Bundesministerium f\"{u}r Bildung und Forschung/Deutsches Zentrum f\"{u}r
Luft- und Raumfahrt (BMFT/DLR), the Max-Planck Society and the
Heidenhain-Stiftung, and also by PPARC, CEA, CNES, and ASI. AF acknowledges
receiving the Max-Plank-Gesellschaft Fellowship. Authors thank Konrad
Dennerl for applying the experimental bias correction software to the
Coma-11 observation.  AF profited from discussions with Michael Freyberg,
Margaret Geller, Fabrizio Nicastro, Maxim Markevitch, Monique Arnaud, Doris
Neumann, Alexey Vikhlinin, and Francesco Miniati. AF would like to
particularly mention a continuous support of this work by Richard Lieu and
his collaborators, Max Bonamente, John Mittaz, and Jelle Kaastra. JPH thanks
MPE for its hospitality during several extended visits. We acknowledge an
anonymous referee for pointing out an importance of the detailed comparison
between XMM-Newton and RASS.
\end{acknowledgements}

\bibliographystyle{aabib99}

\end{document}